\documentclass[pdflatex,sn-nature]{sn-jnl}

\usepackage{graphicx}%
\usepackage{multirow}%
\usepackage{amsmath,amssymb,amsfonts}%
\usepackage{amsthm}%
\usepackage{mathrsfs}%
\usepackage[title]{appendix}%
\usepackage{xcolor}%
\usepackage{textcomp}%
\usepackage{manyfoot}%
\usepackage{booktabs}%
\usepackage{algorithm}%
\usepackage{algorithmicx}%
\usepackage{algpseudocode}%
\usepackage{listings}%
\usepackage{longtable}  
\usepackage{caption}

\definecolor{darkblue}{RGB}{0, 70, 140}   
\definecolor{darkpurple}{RGB}{120, 50, 150} 

\lstdefinestyle{yaml}{
	basicstyle=\color{darkpurple}\footnotesize,
	rulecolor=\color{darkblue},
	string=[s]{'}{'},
	stringstyle=\color{darkblue},
	comment=[l]{:},
	commentstyle=\color{darkblue},
	morecomment=[l]{-}r,
            breaklines = true
}
\lstdefinestyle{yaml-file}{
    breaklines=true,
	basicstyle=\color{darkpurple}\scriptsize,
	rulecolor=\color{darkblue},
	string=[s]{'}{'},
	stringstyle=\color{darkblue},
	comment=[l]{:},
	commentstyle=\color{darkblue},
	morecomment=[l]{-}r,
    numbers=none,
}

\theoremstyle{thmstyleone}%

\theoremstyle{thmstyletwo}%

\theoremstyle{thmstylethree}%

\begin{document}

\title{FHIRconnect: Towards a seamless integration of openEHR and FHIR}


\author*[1,2,5,4]{\fnm{Severin} \sur{Kohler}}\email{severin.kohler@bih-charite.de}

\author[7,8]{\fnm{Jordi} \sur{Piera-Jiménez}}

\author[3]{\fnm{Michael} \sur{Anywar}}

\author[9,10]{\fnm{Lars} \sur{Fuhrmann}}

\author[13,14]{\fnm{Heather} \sur{Leslie}}

\author[1]{\fnm{Maximilian} \sur{Meixner}}

\author[1,12]{\fnm{Julian} \sur{Saß}}

\author[4]{\fnm{Florian} \sur{Kärcher}}

\author[6]{\fnm{Diego} \sur{Boscá}}

\author[11]{\fnm{Birger} \sur{Haarbrandt}}

\author[5]{\fnm{Michael} \sur{Marschollek}}

\author*[1,2,4,15]{\fnm{Roland} \sur{Eils}}\email{roland.eils@bih-charite.de}

\affil*[1]{\orgdiv{Digital Health Center}, \orgname{Berlin Institute of Health at Charité - Universitätsmedizin Berlin}, \orgaddress{\city{Berlin}, \country{Germany}}}

\affil[2]{\orgname{Freie Universität Berlin}, \orgaddress{\city{Berlin}, \country{Germany}}}

\affil[3]{\orgdiv{Department of Health Technologies}, \orgname{Tallinn University of Technology}, \orgaddress{\city{Tallinn}, \country{Estonia}}}

\affil[4]{\orgdiv{Health Data Science Unit}, \orgname{Heidelberg University Hospital and BioQuant}, \orgaddress{\city{Heidelberg}, \country{Germany}}}

\affil[5]{\orgname{Peter L. Reichertz Institute for Medical Informatics of TU Braunschweig and Hannover Medical School}, \orgaddress{\city{Hannover}, \country{Germany}}}

\affil[6]{\orgname{VeraTech for Health}, \orgaddress{\city{Valencia}, \country{Spain}}}

\affil[7]{\orgdiv{Digitalization for the Sustainability of the Healthcare System (DS3) Research Group, Bellvitge Biomedical Research Institute}, \orgaddress{\city{L’Hospitalet de Llobregat}, \country{Spain}}}

\affil[8]{\orgname{Catalan Health Service}, \orgaddress{\city{Barcelona}, \country{Spain}}}

\affil[9]{\orgdiv{Department of Ophthalmology}, \orgname{Asklepios Clinic North - Heidberg}, \orgaddress{\city{Hamburg}, \country{Germany}}}

\affil[10]{\orgdiv{Department of Ophthalmology}, \orgname{University Hospital Düsseldorf}, \orgaddress{\city{Düsseldorf}, \country{Germany}}}

\affil[11]{\orgname{Vitagroup AG}, \orgaddress{\city{Mannheim}, \country{Germany}}}

\affil[12]{\orgname{HL7 Deutschland}, \orgaddress{\city{Berlin}, \country{Germany}}}

\affil[13]{\orgname{Atomica Informatics}, \orgaddress{\city{Melbourne}, \country{Australia}}}

\affil[14]{\orgname{openEHR International}, \orgaddress{\city{Cardiff}, \country{UK}}}

\affil[15]{\orgname{Intelligent Medicine Institute}, \orgdiv{Fudan University}, \orgaddress{\street{131 Dongan Road}, \city{Shanghai}, \postcode{200032}, \country{China}}}

\abstract{
Healthcare interoperability between openEHR and HL7 FHIR remains challenging due to fundamental differences in data modeling approaches and the lack of standardized transformation mechanisms. This paper presents FHIRconnect, a novel Domain-Specific Language (DSL) and open-source transformation engine that enables standardized, bidirectional data exchange between openEHR and FHIR healthcare standards. Our approach addresses critical interoperability gaps through a triple-layered architecture. This design enables 65\% mapping reuse across projects through international archetype-based foundations while accommodating local customizations. FHIRconnect successfully mapped 24 international archetypes to 15 FHIR profiles across seven clinical domains. Key contributions include the first comprehensive DSL for openEHR-FHIR transformation with formal specification, an open-source execution engine (openFHIR) with an accessible mapping library covering high-impact clinical archetypes. This work establishes the technical foundation for community-driven mapping standardization, reducing the need for custom ETL solutions while advancing syntactic and semantic interoperability in healthcare IT system based on open standards.}


\maketitle

\section*{Graphical abstract}
\begin{figure}[h]
	\centering
	\includegraphics[width=.8\textwidth]{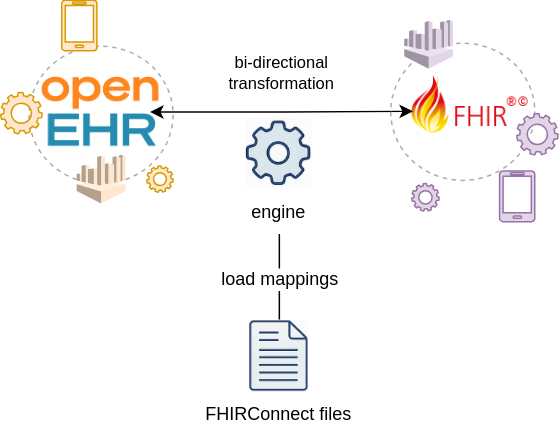}
\end{figure}
\newpage

\section{Introduction}
Healthcare interoperability represents one of the most persistent and costly challenges facing modern health information systems. Despite widespread adoption of Electronic Health Records (EHRs) and substantial investments in health information technology, semantic and structural interoperability across healthcare systems remains largely unrealized \cite{de2022semantic} \cite{palojoki_semantic_2024}. Semantic interoperability ensures that exchanged clinical data maintain consistent meaning across different systems and contexts. Syntactic interoperability requires standardized data formats and organizational schemas that enable automated processing without human interpretation. The absence of both creates profound operational and clinical challenges  \cite{de2022semantic} \cite{Commission2013} \cite{dentler2013barriers}. These challenges have significant consequences for clinical care and research.

The scale of the problem is substantial, but so is the opportunity. The European Health Data Space (EHDS) regulation estimates that transforming digital healthcare could save €11 billion in savings over the next decade by enhancing data accessibility alone. Furthermore, effective interoperability is expected to improve healthcare service efficiency across EU member states, to drive a 20-30\% expansion in the digital health sector while strengthening policy development and scientific research, and to ultimately lead to better health outcomes for European citizens \cite{EHDS}.

Yet, current reality falls far short of this potential. Patient data remain fragmented across multiple systems and institutions, each employing distinct data structures, terminologies, and software platforms \cite{de2022semantic}. Even when identical EHR products are deployed across healthcare sites, configuration variability and implementation differences create semantic inconsistencies that significantly hinder interoperability \cite{bernstam_quantitating_2022}.

As a result, clinical data are frequently siloed and difficult to reuse, obstructing efforts in clinical decision support, quality monitoring, and secondary research \cite{dentler2013barriers}. Healthcare organizations continue to invest heavily in custom Extract, Transform, Load (ETL) pipelines \cite{Gaglova2020, wang2024clinical, cheng2022etl} to move data between systems. These are inherently complex, severely impacted by upstream data source changes, and thus require architectural design and automation to be maintainable \cite{Kimball2013}. Data silos obstruct the  ability of health professionals to access patient information, undermining the quality of care, compromising clinical decision-making, and limiting both quality monitoring initiatives and research capabilities \cite{Commission2013} \cite{ranchal2020} \cite{dentler2013barriers}. The inability to seamlessly exchange and analyze health data across systems represents a fundamental barrier to realizing the transformative potential that the EHDS regulation envisions.

Evidence demonstrates that effective interoperability delivers substantial value. Studies show that improving interoperability enhances data availability, promote safe and effective care, and supports clinical efficiency \cite{li2022impact, Commission2013}. The EHDS framework itself represents recognition of this strategic importance, establishing both the legal foundation and economic justification for prioritizing interoperability through policy frameworks and technical implementation standards \cite{EHDS}.

To enable interoperability, the health informatics domain has witnessed the emergence of multiple standards, most notably openEHR, Fast Healthcare Interoperability Resources (FHIR), and Observational Medical Outcomes Partnership (OMOP) Common Data Model (CDM). Each standard is designed for specific use cases and operates at different layers of the digital health infrastructure \cite{Tsafnat2024}. These standards are not natively interoperable. Their conceptual foundations, structural granularity, and target applications differ, making it difficult to integrate them seamlessly within a single infrastructure \cite{pedrera-jimenez_can_2023}.

Achieving interoperability thus demands carefully designed, semantically aligned mapping processes that handle disparities in data structures, value sets, and clinical meaning. These processes must ensure accurate and contextually valid transformations \cite{Commission2013}. In response to the fragmentation caused by heterogeneous health information standards, research has focused on developing ETL processes between openEHR, FHIR, and OMOP to support semantic integration and data reuse \cite{kohler_eos_2023,ladas2022openehr,PENG2023}. Bridging the semantic and structural gaps between these standards has become a central focus of recent efforts to achieve interoperability.

To achieve interoperability, data must be standardized. Given that most EHRs are not standardized, healthcare organizations predominantly depend on ETL pipelines to facilitate data integration. These pipelines are inherently complex, demanding substantial resources to implement and maintain \cite{Kimball2013}. Moreover, institutions often develop redundant ETL processes across departments, resulting in semantic inconsistencies and data quality issues that compromise the reliability of downstream analyses. The absence of a unified data standard further exacerbates maintenance challenges, with the operational burden escalating as independent workflows continue to proliferate \cite{cheng2022etl}.

Within this context, many organizations seek more sustainable and standardized approaches to streamline data transformation across systems. Once a data-set is standardized, automatic transformations between standards have proven to be far more efficient than maintaining separate bespoke ETL workflows from unstandardised heterogeneous source data for each standard \cite{kohler_eos_2023, haarbrandt_automated_2016, PENG2023}. Recent research has proposed using openEHR as a central integration layer for such healthcare data standards transformation to address these ETL limitations. The semantically rich international data models of openEHR enable comprehensive mappings to other standards, which are shareable between institutions. As a result, this fosters scalable, reproducible, and interoperable clinical data, lowering the costs and complexity of ETLs making more data readily accessible \cite{kohler_eos_2023, haarbrandt_automated_2016, pedrera-jimenez_can_2023, biermann2025}.

We have already defined such an automatic transformation for openEHR-to-OMOP \cite{kohler_eos_2023}. While there are several papers addressing the need for a transformation between openEHR and FHIR \cite{Tsafnat2024, pedrera-jimenez_can_2023, Haarbrandt2018, meredith_aligning_2023, scheuermann_open_2024} and projects to extract data from openEHR and FHIR \cite{bonisch2022harvesting}, only a few address the actual transformation of openEHR-to-FHIR. Rajput et al. defined a context-specific mapping table \cite{rajput2021mapping}, while Ladas et al. developed a context-specific uni-directional mapping tool that transforms two openEHR compositions \cite{ladas2022openehr}. Both approaches are highly context-specific and cannot be generalized across different clinical domains or implementation scenarios. Moreover, no available research provides comprehensive tooling or reusable mappings of openEHR-to-FHIR that could support broader interoperability efforts. 

The core objective of this work is to design, implement, and evaluate a robust, open, and future-proof framework for achieving automated, bidirectional interoperability between openEHR and HL7 FHIR systems. The development of such a framework would promote consistency, transparency, and efficiency in the exchange of clinical data, making it readily usable for both primary care delivery and secondary purposes. This would enhance the interoperability of health information infrastructures in line with the vision set out in initiatives such as the EHDS regulation.

\section{Results}
\begin{figure}[h]
	\centering
    \caption{\textbf{Interaction between FHIRconnect mapping files and the engine}}
	\includegraphics[width=.8\textwidth]{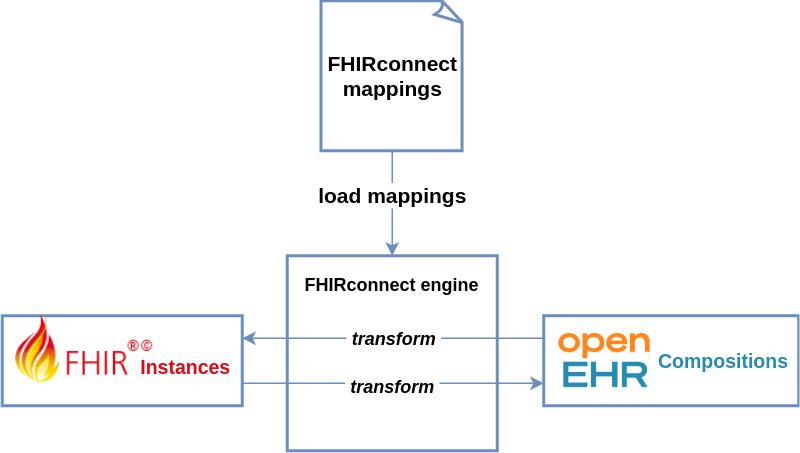}
    \label{engine}
\end{figure}

The Domain-Specific Language (DSL) known as FHIRconnect was developed to support bidirectional mappings between openEHR and FHIR. The mapping files are both human-readable and machine-actionable. These files can be loaded by an engine to enable the automatic transformation of the specified mappings, as illustrated in Figure \ref{engine}. A specification \cite{FHIRconnect}, a library of base mappings \cite{mappingLib}, and an execution engine called openFHIR were implemented. All components are open source and available on GitHub \cite{FHIRconnect} \cite{mappingLib} \cite{openFHIR}.

\subsection{FHIRconnect}
FHIRconnect was implemented to enable bidirectional mapping between openEHR and FHIR. The DSL is based on YAML \cite{BenKiki}, which is designed to be easily authored and read while remaining machine-actionable. The mappings are built around the idea that each path in openEHR and FHIR has a corresponding one in the other standard. As an example, the diagnosis name 'fever' has a path in openEHR and FHIR. This value can be moved either from openEHR into the FHIR path, or from FHIR into openEHR path. As a result, these paths are mapped against each other to allow transformation. An example of a path mapping for the archetype openEHR-EHR-EVALUATION.problem\_diagnosis.v1 \cite{CKM2025} and the FHIR Condition resource \cite{FHIR25} is illustrated in table \ref{chapter_3:mapping}. These mappings are defined based on archetypes and are mapped against resources and their profiles. Templates do not change these archetype paths, making archetype mappings reusable. If based on international archetypes, this provides a sustainable approach for sharing and defining internationally valid mappings. 

\begin{table}[h]
	\caption{\textbf{Example path mappings for problem diagnosis}}
	\begin{tabular}{ccc}
		\hline
		Name  & openEHR path &  FHIRpath \\
		\hline
		Problem/Diagnosis name& data[at0001]/items[at0002] & code \\
		
		Date/time of onset & data[at0001]/items[at0077] & onsetDateTime \\
		\hline
	\end{tabular}
	\centering
	\label{chapter_3:mapping}
\end{table}

\subsection{A multi-layered approach}
\begin{figure}[h]
	\centering
    \caption{\textbf{Overview of the different mapping file types defined in FHIRconnect.}}
	\includegraphics[width=.5\textwidth]{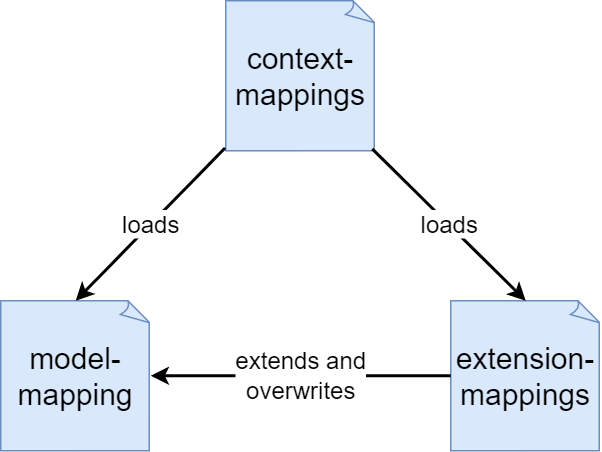}
    \label{fileTypes}
\end{figure}

 To incorporate these different characterizations of resources through profiling in FHIR, FHIRconnect is made up of different layers illustrated in figure \ref{fileTypes}. The model-mappings act as a base layer, defining the transformation between archetypes and resources. The extension-mappings cover FHIR profiles and how archetypes are nested within the specific template. The context-mappings define which template and profile is transformed and imports model-mappings and extensions-mappings.

The first and basic layer is the model-mappings. These describe generic transformations between openEHR archetypes and FHIR resources. For each archetype, a transformation into one or several FHIR resources is defined. Each archetype node is mapped to a resource path if possible. Since openEHR archetypes and resources are used as base models, this provides a generic library of internationally valid mappings for projects. Apart from leveraging reuse and lowering costs, this also defines how fields are transformed between the standards, improving consistency between data due to standardized mappings. An example of such an archetype-to-resource mapping is illustrated in Figure \ref{fhirconnect_model_mapping}. The initial section between 'grammar' and 'StructureDefinition' defines the header which includes the grammar used, specifies the mapping type, metadata, target standard and version, and the corresponding openEHR archetype and FHIR StructureDefinition. The content of the header is standardized between FHIRconnect and the openEHR-to-OMOP mapping language OMOCL \cite{kohler_eos_2023}. In the \textit{mappings}, the fields of both models are mapped against each other: the diagnosis name and date of onset paths of both standards.

A FHIRconnect engine using this mapping extracts values from the specified openEHR or FHIR path and transforms them to the corresponding output path in the other standard. The transformation of the data contained in the paths requires the mapping of the data-types of both standards. As part of FHIRconnect, we defined how these data-types have to be transformed. Not all data types align seamlessly between openEHR and FHIR, and minor workarounds were necessary. Due to the volume and technical complexity of the data types, these are not included in this paper, but are freely available in the FHIRconnect specification \cite{FHIRconnect}. 

\begin{figure}[h!]
\caption{\textbf{FHIRconnect mapping snippet for problem diagnosis.}}
\begin{lstlisting}[style=yaml]
grammar: FHIRConnect/v1.0.0
type: model
metadata:
  name: EVALUATION.problem_diagnosis.v1
  version: 0.0.1-alpha
spec:
  system: FHIR
  version: R4
  openEhrConfig:
    archetype: openEHR-EHR-EVALUATION.problem_diagnosis.v1
    revision: 1.4.1
  fhirConfig:
    structureDefinition: http://hl7.org/fhir/StructureDefinition/Condition

mappings:    
  - name: "problemDiagnose"
    with:
      fhir: "$resource.code"
      openehr: "$archetype/data[at0001]/items[at0002]"

  - name: "dateTime"
    with:
      fhir: "$resource.onset"
      openehr: "$archetype/data[at0001]/items[at0077]"

\end{lstlisting}
	\caption*{Example snippet of a FHIRconnect mapping between the international archetype 
    openEHR-EHR-EVALUATION.problem\_diagnosis.v1 and the FHIR Condition resource. 
    The original full configuration is provided in Supplementary Listing \ref{full_pb}.}
	\label{fhirconnect_model_mapping}
\end{figure}

In some cases archetype and resource paths cannot be mapped 1-to-1. This may require that specific conditions are met for a mapping to be valid, such as hierarchical fields being at different levels of the model. FHIRconnect introduces various functionalities to address these problems. This includes condition-clauses, hierarchical mappings, iteration over paths, etc. A detailed description of these functionalities can be found in the specification \cite{FHIRconnect}.

The extension-mappings are used to adapt to changes introduced by FHIR profiling and to reflect the nesting of archetypes within templates. In FHIR, profiling can modify resources and add fields via extensions. Extension-mappings extend existing model-mappings to cover these FHIR extensions, making them an archetype-to-profile transformation. They also capture whether the archetype mapped, contains another nested archetype introduced by a template. To support this logic, extension mappings provide the necessary functionality: existing model mappings can be extended by adding new path-to-path mappings or by overwriting and enhancing existing ones with additional logic. The approach provides implementers with a stable base set of model-mappings and the flexibility to adapt them to changes introduced by profiling and templates. Both templates and profiles are done project-driven requiring addition of extension-mappings to address specific changes. Around 35\% of the fields in FHIR implementation guides are extensions \cite{kramer_fhir_2022}. Therefore, it is expected that approximately 35\% of the path-to-path mappings will be handled through extensions. This allows a reuse of around 65\% of mappings between projects, due to the model-mappings. The archetype nesting in templates does not introduce any new fields, therefore they only require to trigger other archetype mappings. 

Both model-mappings and their corresponding extension-mappings are grouped within a context-mapping. A context-mapping defines which specific mappings are required to transform a set of FHIR profiles and resources to an openEHR template. It acts as an inclusion mechanism only, it does not modify the mappings themselves. While the transformation is explicitly defined for the profiles, resources and template, it can indirectly trigger other context-mappings through references in FHIR or links within openEHR. Since the transformation is based on archetypes it is mostly agnostic of the template used to input the data. This allows FHIRconnect to extract data independently of the project that was used to ingest it.

For example, FHIR data from the CDS could be transformed to openEHR and then transformed into the US core profiles \cite{uscore}. This simplifies the transformation process for users across FHIR projects. For transformation from FHIR to openEHR, a corresponding template is required to ingest the data because the input of openEHR data needs to be based on templates defining which archetypes are used.

\subsection{Proof-of-concept}
As a proof-of-concept, the base modules of the CDS in FHIR R4 \cite{Profiles2025} and openEHR \cite{HighmedCKM2025} were mapped: Diagnosis, Medication, Person, Procedure, Laboratory, Case, Consent. The CDS is used by 38 university hospitals in Germany to exchange data for secondary use as part of the German Portal for Medical Research Data (FDPG), standardizing data from over 21 million patients, 250 million diagnosis, and 2 billion laboratory data sets \cite{FDPG}. The mappings were based on the templates mapping all supported fields of the FHIR profiles. For four profiles a corresponding template was missing, therefore these where not mapped, a listing of the mapped profiles can be found in Supplementary table \ref{appendix:profiles}. We transformed sample data successfully with the FHIRconnect mappings and the open source engine \cite{openFHIR}. These mappings are open source and can be found under \cite{mappingLib}. 

This resulted in 12 context-mappings files, 21 extensions-mappings and 37 model-mapping files. The FHIRconnect mappings cover 24 international archetypes, 6  local archetypes and 15 profiles with 34 extensions. The Supplementary Information provides detailed listings of the mapped extensions (table \ref{appendix:extensions}), profiles (table \ref{appendix:profiles}), and archetypes (table \ref{appendix:archetypes}). Several of the local archetypes are marked as legacy and scheduled for deletion. Therefore, they were not included in the count. 

The 24 international archetypes cover around 9.7\% of all published archetypes in the CKM \cite{CKM2025}. Published archetypes are archetypes that went through a content peer review process resulting in consensus that the archetypes are fit and safe for use. Thus, they meet established standards for clinical validity, consistency, and interoperability. Depending on the context of health information, these are widely used. As an example, the problem\_diagnose archetype is used in 112 templates from the international \cite{CKM2025} and German CKM \cite{HighmedCKM2025} (as of 18.06.2025). Other notable mappings are the laboratory\_test\_result.v1, medication.v1, medication\_statement.v0, procedure.v1.  These mappings cover the archetypes typically used in summaries like the International Patient Summary \cite{kay2020international}. Due to the international nature of archetype mappings, these mappings are valid not only for the CDS, but also for all templates using these archetypes. Therefore, they lay the foundation for an international mapping library between openEHR and FHIR. 

\subsection{Comparing openEHR and FHIR}
Due to the generic nature of resources, some of the archetype fields have no mapping in FHIR. As an example, the procedure archetype, which represents the basic concept of a clinical procedure, is missing 8 fields that are not present in the equivalent FHIR resource (see Supplementary Information, table \ref{appendix:procedure}). Given that this represents a core clinical concept, it is notable that at this level certain key fields are already missing. This suggests that for more specific clinical models, such as those dealing with reproductive cycles, there will likely be even greater gaps or missing data. FHIR resources are designed to capture only the information present in most systems, more specific data needs to be modelled via extensions. These extensions, if not internationally harmonized, lead to interoperability problems. Each project can introduce their own type of extensions to cover fields missing. Currently, extensions make up around 35\% of the fields in FHIR projects, four out of five of these extensions are non-standardized and therefore lead to interoperability problems between projects \cite{kramer_fhir_2022}. If a FHIR profile does not cover these archetype fields via extensions, this data will be lost. Data loss from FHIR to openEHR is less likely to occur, as openEHR modeling aims for a maximum-model. In cases where archetypes or fields are missing, they need to be added as part of the international modeling process. Aligning FHIR profiles or generating them directly from archetypes could minimize this loss. 

Some data types in openEHR and FHIR do not align exactly, but were solved via workarounds. In two cases, FHIR data types mappings were not a data value in openEHR, but a cluster. This transformation needs to take these factors into account. Resulting gaps and the need for alignment were reported to the HL7 and openEHR communities, where the topic is currently under discussion. A detailed description of the gaps and workarounds can be found in the specification \cite{FHIRconnect}.

Some archetype node fields have no associated terminology codes. For example, most nodes in the Assisted Reproduction Treatment Cycle Summary archetype \cite{CKM2025} do not have standardized terminology codes. openEHR defines meaning through its structured models, e.g., the node path for Number of oocytes collected uniquely identifies the field, including translations into other languages along with terminology bindings. In contrast, FHIR relies heavily on external terminologies to define and identify clinical concepts, as well as for translations. For example, a blood pressure Observation in FHIR is identified by its specific code from a terminology. Without this code, the Observation cannot be recognized as a blood pressure measurement and would lack clinical meaning. In cases where no standardized terminology codes are available in e.g. SNOMED, LOINC, custom internal codings are required as an ad hoc and less sustainable solution. This highlights a limitation of relying heavily on terminologies to convey model semantics.

\section{Discussion}
A set of FHIRconnect mappings and an open-source engine were successfully implemented and sample data transformed. The results showed that a standardized ETL process between openEHR and FHIR and vice-versa, is feasible, but requires custom extensions to cover more than just resources. This eases the transformation of data between openEHR and FHIR, making data more interoperable between projects. Here, we provide a first set of openEHR-to-FHIR mappings and an open-source tool to execute FHIRconnect mappings. These mappings cover some of the most commonly used archetypes and resources, laying the foundation for an international library of openEHR-to-FHIR mappings. FHIRconnect lowers the barrier to accessing openEHR and FHIR data between projects and makes secondary-use data, e.g. from the FDPG, accessible in both openEHR and FHIR.

FHIRconnect defines a human-readable and machine-actionable declarative language to map openEHR and FHIR bidirectionally. The language has the capability to define all mappings necessary to represent the CDS. We implemented an open-source solution, openFHIR, to use these mappings for data transformation. Ultimately, this makes all the data contained in the mapped archetypes transformable to the CDS, including data from several national records based on openEHR in Europe and the German FHIR CDS data, which currently covers over 20 million patients.

The FHIRconnect mappings cover approximately 9.7\% of all published archetypes in the CKM \cite{CKM2025}. Although the number may seem small, these archetypes are among the most commonly used and encompass the majority of the data needed for simple patient summaries. The objective of this work was to develop a method for integrating both standards. This integration is made possible through FHIRconnect. We hope that the provided mappings are the beginning of a cross-community initiative to build an international mapping library between openEHR and FHIR, which will eventually cover the remaining archetype and resource mappings.

FHIR projects face the challenge that approximately 35\% of the contained data must be modelled using extensions. This leads to interoperability issues between projects, as other institutions are unable to interpret or map these fields ad hoc. Instead, they require additional information, such as that provided by an FHIR Implementation Guide (IG), which defines and describes how FHIR resources are used in a specific context through constraints, profiles, and terminology bindings to support interoperability. These fields often result in duplicated extensions across projects, since they can be added or modelled differently. Consequently, users must transform their data storage each time they encounter such an extension in order to harmonize these fields. This also has drawbacks for analytics and persistence in FHIR, as semantically equivalent data may be stored in different fields.

Using openEHR as a persistence layer or as the basis for clinical requirements for FHIR data, offers a solution to this extension integration problem. Archetypes are expected to cover these extensions, if not, missing fields are typically added, due to openEHR’s inclusive modeling approach. This provides users with an integration layer that is independent of the specific use-cases of any individual FHIR project, facilitating both integration and data extraction. Since mappings are defined on tightly governed archetypes, this also enables ingesting and transforming data across different FHIR projects. As an example, the CDS data could be transformed into openEHR and from there transformed into the Sparked \cite{AUCDI} profiles, if mappings were provided. Moreover, openEHR is not affected by the six breaking versions of FHIR (R1-R6), making it a more stable source for integration. Thus, openEHR can serve as a reliable data source for FHIR data, thus synergistically making use of the specific advantages of both standards. FHIRconnect, as an implementation, enables the necessary transformation between the openEHR persistence layer and the FHIR exchange layer.

When openEHR is used as the basis for deriving FHIR profiles directly from archetypes, it has the potential to significantly improve interoperability among FHIR projects. This approach requires the systematic inclusion of FHIR extensions that correspond to the underlying archetype structures. Establishing a standardized methodology for defining and placing these extensions within FHIR profiles could ensure consistent and semantically interoperable representations across different FHIR implementations effectively addressing the issue of FHIR "proliferation". In this context, model mappings could support standardized extensions directly, enabling their automatic instantiation and integration into transformation processes.

This approach is already being implemented in projects such as the Sparked project, where FHIR profiles and IGs are modelled in reference to openEHR archetypes \cite{AUCDI}. It also simplifies the creation of FHIRconnect mappings. The more FHIR profiles deviate from the hierarchical structure and design of archetypes, the more complex the mapping becomes. While FHIRconnect provides the necessary tooling, the overall complexity of the mapping process is heavily dependent on the alignment of the underlying models. Poorly aligned models reduce the number of reusable mappings and require additional effort to maintain compatibility. If data is already modelled using openEHR, it can be more efficient to use openEHR directly for exchange, avoiding mapping complexity and potential data loss. Furthermore, openEHR is designed to standardize data at the point-of-capture effectively preventing expensive retrospective standardization of systems capturing data in a non-standardized way into e.g. FHIR. openEHR defines separate models for EHR and demographic information and does not mandate the use of its demographic model. Some implementations combine openEHR for EHR records with a FHIR demographics server \cite{BetterFHIR}. 

This heterogeneity reinforces the need for standardized, bidirectional transformation between the two standards. FHIR is widely adopted across healthcare, the here presented tool FHIRconnect plays an essential role in enabling interoperability across different healthcare systems.

\subsection{Limitations}
Despite its promising capabilities, FHIRconnect currently covers only a subset of archetypes and profiles, which highlights the need for ongoing community efforts to expand the mapping library. 
Apart from that, not all data types and values in FHIR and openEHR could be directly transformed and required workarounds. The resulting findings were reported to the HL7 and openEHR communities and are under discussion. Additionally, mapping complexity grows as FHIR profiles diverge from archetype structures, potentially requiring customization. Future work will focus on broadening archetype coverage, improving automation in mapping generation and validating transformations across diverse real-world datasets. This will help to ensure data integrity, reduce loss during transformation, and facilitate wider adoption in clinical and research settings.

\subsection{Conclusion}
This work successfully implemented a standardized, bidirectional transformation language between openEHR and FHIR along with an open-source engine. By defining reusable, archetype-based mappings, FHIRconnect enables more consistent and interoperable data exchange across projects using either standard. The proof-of-concept mapping of the CDS validates the practical applicability for real-world healthcare data. This represents the first implementation of a comprehensive openEHR to FHIR bidirectional transformation language with engine and thus represents a significant step towards achieving interoperability between the two widely used standards. 

While it is possible to transform data from openEHR to FHIR, the process requires extensive customization through FHIR extensions. Projects that do not align with openEHR archetype models risk losing some of the structured clinical data during transformation. For organizations not using openEHR, adopting archetype-based profiles within FHIR can significantly improve semantic consistency and data quality.
However, these efforts often replicate capabilities already built into openEHR and adopting openEHR directly can save resources and ensure higher semantic precision from the outset. Where needed, data can still be mapped to FHIR for external interoperability and exchange scenarios. Tools like FHIRconnect facilitate this integration. This approach ultimately advances the broader goal of making high-quality health data more interoperable, portable, and semantically consistent across diverse healthcare systems.

\section{Methods}

\subsection{Study Design and Approach}
This study employed a design science research methodology to develop and validate a domain-specific language for bidirectional data transformation between openEHR and FHIR. The research approach consisted of four main phases: (1) analysis of existing standards and mapping requirements, (2) design and implementation of the FHIRconnect DSL and execution engine, (3) proof-of-concept validation using the german core dataset (CDS), and (4) evaluation of mapping coverage and transformation accuracy.

\subsection{FHIR}
The HL7 FHIR standard is a widely adopted framework for healthcare data exchange that combines a standardized data model with a modern API-based exchange mechanism. At its core, FHIR defines resources, which are modular, reusable data structures that represent clinical and administrative concepts. These resources are exchanged using RESTful APIs and other web technologies. Each resource represents information about a given scope, for example, a patient. FHIR provides a formal mechanism for adapting its core resource models to specific use cases through a process called profiling. While the FHIR core specification defines broad and flexible resource structures, real-world implementations often require more precise definitions. Profiling allows implementers to constrain fields, define fixed values, specify terminology bindings, and introduce new fields through extensions \cite{FHIR25}. To ensure consistency and interoperability for specific projects, sets of these  resources and profiles are combined into a FHIR Implementation Guide (IG), which provides a detailed specification on how these models should be used in a particular context, including constraints, and terminology bindings. FHIR can be adapted by vendors with a proprietary database model by introducing a FHIR facade which maps the internal model the API, making the standard adaptable across diverse systems. However, current research has shown that FHIR resources cover approximately 65\% of required information needs \cite{kramer_fhir_2022}. The remaining 35\% often necessitates the use of extensions, the majority of which are custom-defined and not part of the FHIR core specification or the official FHIR Extensions Pack. This proliferation of implementation specific extensions introduces new interoperability challenges.

\subsection{openEHR}
openEHR is designed to support semantically rich and structured documentation of clinical data through archetype-based modeling \cite{Beale2002, Archetype2025}. The openEHR specification defines an interoperable architecture to standardize EHR systems using a multi-layered information model approach. The base layer consists of the Reference Model (RM) which defines basic clinical concepts and attributes \cite{RM2025}. The second layer comprises reusable models known as archetypes, which are community-defined and represent detailed clinical models, such as heart rate measurements, procedures or diagnosis. These archetypes aim to include data fields that support all intended uses and contexts for the clinical information they represent. openEHR excels when used at the point-of-capture for longitudinal clinical documentation, though this comes with high initial adaptation costs \cite{Archetype2025}.

Archetypes encompass a broad range of information, often including more than necessary for specific applications, as they are designed as maximum dataset models. To represent specific clinical use-cases, templates are used. They represent the third tier of models and use archetypes as building blocks. Archetypes are developed and maintained collaboratively by the openEHR community as internationally shared models, published in the Clinical Knowledge Manager (CKM) the official openEHR international archetype library \cite{CKM2025}. If required, archetypes can also be created locally to meet specific requirements, but if not internationally consented this can lead to interoperability problems.

Each node within an archetype is given a unique archetype node identifier. When archetypes are incorporated into a template, these identifiers, together with the archetype ID, are preserved. They are then used to create paths, which allow data to be accurately retrieved from archetype nodes embedded within templates. 

These paths are used to retrieve data from archetype nodes embedded within templates. Templates cannot add new fields to archetypes, they only apply constraints to existing archetype fields and define how archetypes are combined and nested to represent specific clinical use cases. Therefore, interoperability is ensured between templates which are using the same archetypes. Template models can be developed for local implementations or even standardized for national use. 

\subsection{Mapping openEHR and FHIR}
Both openEHR and FHIR have clinical models that contain data fields for a given scope. FHIR starts from generic base resource definitions, which can be constrained and extended through profiling to meet use-case requirements. openEHR, follows a maximum data-model approach and constrains out fields at the template level. A mapping between both standards, requires a processable rule-set, which enables the conversion of both the content and the relationships of data fields from one standard to the other while preserving the integrity of their semantic content. 

One approach is to represent an openEHR template as a FHIR Questionnaire, introducing a non-openEHR way of representing the data, moving it closer to FHIR. This approach has the potential to simplify the mapping process by allowing template fields to be “flattened” into a list structure that fits into FHIR Questionnaire items. While this can facilitate simple, rule-based mappings, it comes with significant drawbacks.

A distinct Questionnaire and its associated mapping logic must be generated for each individual template, which reduces scalability. Flattening the multi-level structure of openEHR may result in the loss of semantic and structural information. It also limits the potential for reuse of mappings between projects and institutions, since the underlying archetypes within the template are flattened as well, reducing the opportunity to share and reuse mappings. Questionnaires are "a structured set of questions intended to guide the collection of answers from end-users." \cite{FHIR25}. Therefore, the contained openEHR data still has to be transformed into FHIR resources, requiring an additional layer of transformation logic. The additional transformation steps, potential loss of structural richness and reusability of mappings can significantly increase the effort required for implementation, ongoing maintenance, and long-term sustainability.

Another approach is to build up mappings directly based on the data. This has several potentials. First of all, no pre-processing of data is necessary, it is mapped based on its native format, preventing loss. The mappings defined can be designed in a way to make use of the paradigms and structures of both standards. Promising a high reuse between mappings and lowering maintainment and implementation costs. Therefore, we decided to follow the latter approach, which has proven to be effective in one of our earlier studies \cite{kohler_eos_2023}.

To take into account the information model structure of openEHR and FHIR, the corresponding models have to be mapped against each other. openEHR archetypes are mapped to FHIR resources and profiles but can also represent information that may be found in more than one resource or profile. Archetypes fields need to be processed in a standardized way to create a sustainable mapping. In FHIR this is also true for resources. On the other hand, FHIR profiles can introduce new fields to resources. Therefore, a mapping needs to cover both resources and their extensions.

\begin{figure}[h!]
	\centering
    \caption{\textbf{Steps of an openEHR-to-FHIR mapping}}
	\includegraphics[width=.6\textwidth]{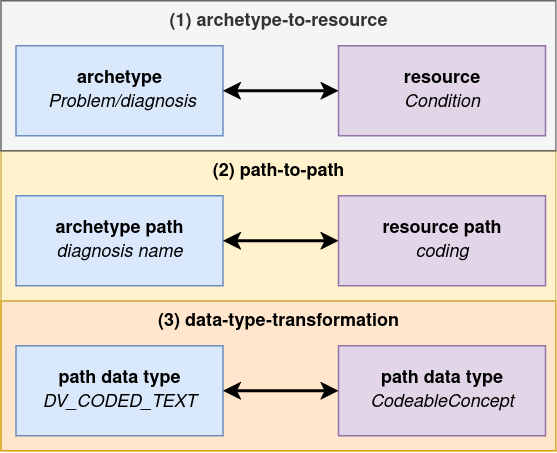}
    \caption*{\textit{Example of the openEHR-to-FHIR mapping steps using the problem diagnosis. (1) the archetype and resource are mapped as a first step of the mapping. (2) each field of the resource and archetype is mapped against each other. (3) the data types of each of these fields need to be transformed to transfer data into another standard}}
    \label{mappingSteps}
\end{figure}    

A mapping can be divided into three sequential steps, illustrated in figure \ref{mappingSteps}. First, the archetype and resource have to be identified and mapped. Secondly, each field of the resource and archetype needs to be mapped as a path-to-path mapping. Finally, this requires that the underlying data type definitions of both openEHR and FHIR, such as coded value or quantity, are automatically transformed into each other.

\backmatter

\section*{Code Availability}
The specification, base mapping library, and the openFHIR execution engine are open source and available on GitHub at \href{https://github.com/SevKohler/FHIRconnect-spec}{FHIRconnect-spec}, 
\href{https://github.com/SevKohler/FHIRconnect-mapping-lib}{FHIRconnect-mapping-lib}, and 
\href{https://github.com/medblocks/openFHIR}{openFHIR}. 
All components are released under the Apache 2.0 license.

\bibliography{mybib}

\section*{Competing Interests}
B.H. is employed by a company that develops and maintains openEHR platforms. H.L. is employed by openEHR International. M.A., D.B., H.L., and S.K. report consultancy work. The remaining authors declare no competing interests.

\section*{Acknowledgements}
The authors thank the HiGHmed association, Medblocks and Gasper Andrejc for the implementation of the open-source engine openFHIR used in this work. We would also like to thank Better specifically Matija Poljnar and Jake Smolka, for contributing to early prototype concepts of FHIRconnect. We thank Dirk Meyer zu Büschenfelde of the Berlin Institute of Health for his insights.

\section*{Author contributions}
S.K. was the primary author of the manuscript, responsible for conceptualization, methodology, software development, validation, and data curation. D.B., F.K., and S.K. contributed to the conceptual design of the specification. M.A., D.B., J.S., H.L., M.M., F.K., B.H., M.Ma., and R.E. contributed to methodology development, validation, or review of the manuscript. J.P.-J., L.F., J.S., H.L., B.H., and M.Ma. critically reviewed and edited the manuscript. R.E. provided supervision.

\bigskip

\section*{Supplementary information}

\begin{lstlisting}[style=yaml-file,
  caption={FHIRconnect mapping of the Problem Diagnosis},
  label={full_pb}
]
grammar: FHIRConnect/v1.0.0
type: model
metadata:
  name: EVALUATION.problem_diagnosis.v1
  version: 0.0.1-alpha
spec:
  system: FHIR
  version: R4
  openEhrConfig:
    archetype: openEHR-EHR-EVALUATION.problem_diagnosis.v1
    revision: 1.4.1
  fhirConfig:
    structureDefinition: http://hl7.org/fhir/StructureDefinition/Condition

preprocessor:
  fhirCondition:
      targetRoot: "$resource.verificationStatus"
      targetAttribute: "coding.code"
      operator: "not of"
      criteria: "entered-in-error"

mappings:

  - name: "contextStartTime"
    with:
      fhir: "$resource.recordedDate"
      openehr: "$composition/context/start_time"

  - name: "participations"
    with:
      fhir: "$resource.asserter"
      openehr: "$composition/context/participations"  
      type: "NONE"
    followedBy:
      mappings:
        - name : "participationFunction"
          with:
            fhir: "$fhirRoot"
            openehr: "$openEHRRoot"
          manual:
            - name: "function"
              openehr:
                - path: "function"
                  value: "asserter"
        - name: "performer"
          with:
            fhir: "$fhirRoot"
            openehr: "$openEHRRoot"
  - name: "composer"
    with:
      fhir: "$resource.recorder"
      openehr: "$composition/composer"


  - name: "other_participations"
    with:
      fhir: "$resource.asserter"
      openehr: "$archetype/diagnose/other_participations"  
      type: "NONE"
    followedBy:
      mappings:
        - name : "participationFunction"
          with:
            fhir: "$fhirRoot"
            openehr: "$openEHRRoot"
          manual:
            - name: "function"
              openehr:
                - path: "function"
                  value: "asserter"
        - name: "performer"
          with:
            fhir: "$fhirRoot"
            openehr: "$openEHRRoot" 

  - name: "provider"
    with:
      fhir: "$resource.recorder"
      openehr: "$archetype/provider"  
      
  - name: "problemDiagnose"
    with:
      fhir: "$resource.code"
      openehr: "$archetype/data[at0001]/items[at0002]"

  - name: "note"
    with:
      fhir: "$resource.note.text"
      openehr: "$archetype/data[at0001]/items[at0069]"

  - name: "dateTime"
    with:
      fhir: "$resource.onset"
      openehr: "$archetype/data[at0001]/items[at0077]"

  - name: "bodySite"
    with:
      fhir: "$resource.bodySite"
      openehr: "$archetype/data[at0012]"

  - name: "bodySiteCluster"
    with:
      fhir: "$resource.bodySite"
      openehr: "$archetype/data[at0001]/items[openEHR-EHR-CLUSTER.anatomical_location.v1]"
    slotArchetype: "CLUSTER.anatomical_location.v1"

  - name: "severity"
    with:
      fhir: "$resource.severity"
      openehr: "$archetype/data[at0001]/items[at0005]"
\end{lstlisting}

\begin{table}[h]
    \caption{Collection of all extensions contained in the mappings of the core-dataset aligned under their modules \cite{Profiles2025} mappings.}
    \centering
    \begin{tabular}{ll}
        \hline
        \textbf{Case} & \textbf{Consent} \\
        \hline
        Aufnahmegrund & domainReference \\
        ErsteUndZweiteStelle & domain \\
        DritteStelle & xacml \\
        VierteStelle & status \\
        plannedStartDate & \\
        plannedEndDate & \\
        Entlassungsgrund & \\
        \hline
        \textbf{Diagnose} & \textbf{Laboratory} \\
        \hline
        ReferenzPrimaerdiagnose & QuelleKlinischesBezugsdatum \\
        Feststellungsdatum & pqTranslation \\
        Seitenlokalisation & quantityPrecision \\
        Diagnosesicherheit & \\
        lebensphase-bis & \\
        lebensphase-von & \\
        \hline
        \textbf{Medication} & \textbf{Person} \\
        \hline
        Wirkstofftyp & other-amtlich \\
        Wirkstoffrelation & Stadtteil \\
         & data-absent-reason \\
         & namenszusatz \\
         & nachname \\
         & vorsatzwort \\
         & prefix-qualifier \\
         & gemeindeschluessel \\
         & Postfach \\
        \hline
        \textbf{Procedure} & \\
        \hline
        Seitenlokalisation & \\
        durchfuehrungsabsicht & \\
        Dokumentationsdatum & \\
    \end{tabular}
    \label{appendix:extensions}
\end{table}

\begin{table}[h]
    \caption{Collection of the core-dataset profiles \cite{Profiles2025} and their modules and if these where mapped. X marks if a profile was mapped. If not mapped a reasons is provided. }
    \centering
    \begin{tabular}{lll}
        \hline
        \textbf{Module} & \textbf{Profile} & \textbf{Mapped} \\
        \hline
        Case &  MII\_PR\_Fall\_Kontakt\_mit\_einer\_Gesundheitseinrichtung   & x \\
        Consent & MII\_PR\_Consent\_Einwilligung  & x \\
         & MII\_PR\_Consent\_DocumentReference & template missing \\
         & MII\_PR\_Consent\_Provenance & template missing \\
        Diagnose & MII\_PR\_Diagnose\_Condition & x \\
        Laboratory & MII\_PR\_Labor\_Laboranforderung & x \\
         & MII\_PR\_Labor\_Laboruntersuchung & x \\
         & MII\_PR\_Labor\_Laborbefund & x \\
        Medication & MII\_PR\_Medikation\_Medication & x \\
         & MII\_PR\_Medikation\_Medikationsliste  &  template missing \\
         & MII\_PR\_Medikation\_MedicationStatement & x \\
         & MII\_PR\_Medikation\_MedicationRequest & x \\
         & MII\_PR\_Medikation\_MedicationAdministration & x \\
        Person & MII PR Person Patient (Pseudonymisiert) & x \\
        &  MII PR Person Vitalstatus  & x \\
        &  MII PR Person Todesursache  & x \\
        & MII PR Person Proband (Pseudonymisiert) &  template missing \\
        &  MII PR Person Patient  (Pseudonymisiert) & x \\
        Procedure &   MII PR Prozedur Procedure  & x\\
        \hline
    \end{tabular}
    \label{appendix:profiles}
\end{table}

\begin{table}[h]
    \caption{List of all archetypes mapped from the CKM \cite{HighmedCKM2025}, including legacy archetypes.}
    \centering
    \begin{tabular}{p{0.9\linewidth}}
        \hline
        \textbf{German archetypes}  \\
        \hline
        openEHR-EHR-CLUSTER.lebensphase.v0 \\
        openEHR-EHR-ADMIN\_ENTRY.versicherungsinformationen.v0 \\
        openEHR-EHR-ADMIN\_ENTRY.episode\_institution\_local.v0 \\
        openEHR-EHR-ADMIN\_ENTRY.person\_data.v0 \\
        openEHR-EHR-CLUSTER.study\_details.v1 \\
        openEHR-EHR-CLUSTER.study\_participation.v1 \\
        \hline
        \textbf{Legacy german archetypes } \\
        \hline
        openEHR-EHR-CLUSTER.entry\_category.v0 \\
        openEHR-EHR-CLUSTER.multiple\_coding\_icd10gm.v1 \\
        openEHR-EHR-CLUSTER.identifier\_fhir.v0 \\
        openEHR-EHR-CLUSTER.organization.v0 \\
        openEHR-EHR-CLUSTER.medication\_status\_fhir.v0 \\
        openEHR-EHR-CLUSTER.observation\_status\_fhir.v1 \\
        \hline
        \textbf{International Archetypes} \\
        \hline
        openEHR-EHR-ACTION.medication.v1 \\
        openEHR-EHR-ACTION.informed\_consent.v0 \\
        openEHR-EHR-ACTION.procedure.v1 \\
        openEHR-EHR-INSTRUCTION.service\_request.v1 \\
        openEHR-EHR-OBSERVATION.medication\_statement.v0 \\
        openEHR-EHR-OBSERVATION.laboratory\_test\_result.v1 \\
        openEHR-EHR-CLUSTER.person.v1 \\
        openEHR-EHR-CLUSTER.structured\_name.v1 \\
        openEHR-EHR-CLUSTER.person\_birth\_data\_iso.v0 \\
        openEHR-EHR-CLUSTER.medication.v2 \\
        openEHR-EHR-CLUSTER.dosage.v2 \\
        openEHR-EHR-CLUSTER.laboratory\_test\_analyte.v1 \\
        openEHR-EHR-CLUSTER.case\_identification.v0 \\
        openEHR-EHR-CLUSTER.death\_details.v1 \\
        openEHR-EHR-CLUSTER.specimen.v1 \\
        openEHR-EHR-CLUSTER.problem\_qualifier.v2 \\
        openEHR-EHR-CLUSTER.address.v1 \\
        openEHR-EHR-CLUSTER.anatomical\_location.v1 \\
        openEHR-EHR-CLUSTER.organisation.v1 \\
        openEHR-EHR-CLUSTER.timing\_daily.v1 \\
        openEHR-EHR-EVALUATION.vital\_status.v1 \\
        openEHR-EHR-EVALUATION.problem\_diagnosis.v1 \\
        openEHR-EHR-EVALUATION.gender.v1 \\
        openEHR-EHR-EVALUATION.cause\_of\_death.v1 \\
        \hline
    \end{tabular}
    \label{appendix:archetypes}
\end{table}

\begin{longtable}{|p{4cm}|p{4cm}|p{6cm}|}
\caption{Procedure mapping table. A dot in an openEHR path indicates a child element.} \\
\hline
\textbf{openEHR} & \textbf{FHIR} & \textbf{Comment} \\
\hline
\endfirsthead

\hline
\textbf{openEHR} & \textbf{FHIR} & \textbf{Comment} \\
\hline
\endhead

clusters / RM & identifier & \\
\hline
link & instantiatesCanonical & \\
      & instantiatesUri & \\
\hline
link & basedOn & \\
\hline
link & partOf & \\
procedure & partOf (if procedure) & if multiple procedure are allowed in the template\\
\hline
ism\_transition & status & \\
\hline
reason & statusReason & \\
\hline
type & category & \\
\hline
Procedure name & code & \\
\hline
ehr ID & subject & \\
\hline
link & encounter & \\
\hline
composer & recorder & \\
\hline
participations & asserter & \\
\hline
participation.function & performer.function & \\
participation & performer.actor & \\
participation.perfomer & performer.onBehalfOf & Stored in the demographics \\
\hline
event\_context.location & location & \\
\hline
reason & reasonCode & \\
\hline
link & reasonReference & \\
\hline
bodySite & bodySite & \\
\hline
outcome & outcome & \\
\hline
feeder audit.originating system items id & report & \\
\hline
complication & complication & \\
\hline
link & complicationDetail & Link to Condition \\
\hline

comment & note & \\
\hline
Procedure detail.Operative procedure.approach & focalDevice.action & \\
Procedure detail.Operative procedure.medical device & focalDevice.manipulated & maps to an medical device inside operative procedure cluster \\
\hline
Procedure detail & usedReference & \\
\hline
Procedure detail & usedCode & \\
\hline
time  & performedDateTime & \\
Final end date/time &  & \\
start time  &  & if no start time is provided in FHIR, this will be mapped as start since mandatory in openEHR\\
\textbf{NONE} & performedRange & \\
start time & performedPeriod.start &  \\
time & performedPeriod.end & \\
end time &  & \\
Final end date/time &  & \\
\textbf{NONE} & performedString & \\
\textbf{NONE} & performedAge & \\
\textbf{NONE} & performedRange & \\
\hline
Scheduled date/time & \textbf{NONE} &  \\
\hline
\textbf{NONE} & followUp & no direct mapping available due to philosophical differences \\
\hline
Procedural difficulty & \textbf{NONE} & \\
\hline
Urgency & ServiceRequest.priority & \\
\hline
Multimedia & \textbf{NONE} & \\
\hline
Total duration & performedPeriod & could be calculated from performedPeriod\\
\hline
method & \textbf{NONE} & can be covered by the international extensions of FHIR for method \\
\hline
Indication & \textbf{NONE} &  \\
\hline
Description & \textbf{NONE} & could be mapped to note, but would miss contextual information \\
\hline
Scheduled date/time & \textbf{NONE} & \\
\hline
Final date/time & \textbf{NONE} & \\
\hline
Requestor order identification & ServiceRequest & \\
\hline
Requestor & ServiceRequest & \\
\hline
Receiver order identifier & ServiceRequest & \\
\hline
Receiver & ServiceRequest & \\
\hline
instruction details & ServiceRequest &  \\
\hline
\label{appendix:procedure}
\end{longtable}

\end{document}